# Uncovering Obscured Phonon Dynamics from Powder Inelastic Neutron Scattering using Machine Learning

Yaokun Su[1] and Chen Li[1,2*]

1. *Materials Science and Engineering, University of California, Riverside, Riverside, CA, United States.*
2. *Mechanical Engineering, University of California, Riverside, Riverside, CA, United States.*
   *chenli@ucr.edu

## Abstract

The study of phonon dynamics is pivotal for understanding material properties, yet it faces challenges due to the irreversible information loss inherent in powder inelastic neutron scattering spectra and the limitations of traditional analysis methods. In this study, we present a machine learning framework designed to reveal obscured phonon dynamics from powder spectra. Using a variational autoencoder, we obtain a disentangled latent representation of spectra and successfully extract force constants for reconstructing phonon dispersions. Notably, our model demonstrates effective applicability to experimental data even when trained exclusively on physics-based simulations. The fine-tuning with experimental spectra further mitigates issues arising from domain shift. Analysis of latent space underscores the model's versatility and generalizability, affirming its suitability for complex system applications. Furthermore, our framework's two-stage design is promising for developing a universal pre-trained feature extractor. This approach has the potential to revolutionize neutron measurements of phonon dynamics, offering researchers a potent tool to decipher intricate spectra and gain valuable insights into the intrinsic physics of materials.

## Introduction

Phonon dynamics is fundamental for many material properties, such as thermal transport, thermal expansion, lattice structure, and phase transformations. Phonons have significant power in tailoring material properties because of their couplings with electronic and magnetic degrees of freedom[1–5]. Inelastic neutron scattering (INS) is one of the most important tools for phonon measurements[6,7]. The resulting spectrum, often expressed as the dynamical susceptibility $\chi''(\mathbf{Q}, E)$, represents the intensity of scattered neutrons as a function of momentum transfer $\mathbf{Q}$ and energy transfer $E$. Phonon dispersions throughout the entire Brillouin zone are accessible through the scattering between phonons and neutrons. However, four-dimensional measurements are rare due to the strict requirements for large and high-quality single-crystal samples, as well as the scarcity of neutron beam time. For some materials, it is not practical to synthesize single-crystal samples. For some other samples, such as low dimension or nanostructured materials, it is not possible to do so. Only two-dimensional measurements in the $(|\mathbf{Q}|, E)$-space with momentum collapsing into a scalar are possible. While this approach is still useful for characterizing materials, it introduces complexities in extracting full phonon dynamics. The loss of momentum direction-dependent information makes the extraction underdetermined and requires physics-based models. Even with substantial manual efforts in model selection, refinement, and validation, such extraction could be unreliable, and thus was rarely attempted. A consistent and automated approach is urgently needed to extract full phonon dynamics information and gain deeper insights from two-dimensional INS measurements.

With the development of inelastic neutron scattering instruments and the availability of high-performance computing, a data-driven solution to overcome these obstacles is possible. Modern spectrometers such as the Wide Angular-Range Chopper Spectrometer (ARCS)[8] at Oak Ridge National Laboratory has a higher neutron flux and a wider coverage of momentum, enabling more accurate and comprehensive inelastic neutron scattering measurements with better details and less noise than those obtained in the early days of neutron scattering. In addition to advances in instruments, high-performance computing resources have greatly improved the ability to process and analyze large amounts of complex data, replacing the laborious and inefficient manual analysis process. Although optimization methods appear promising for extracting phonon information, the high computational cost of gradient calculations is a significant drawback. Even with the adoption of gradient-free optimization methods, limitations persist, as these approaches require repeating the process for each new spectral analysis. Additionally, optimizing each spectrum individually often leads to overfitting because of the ill-posed nature of this inverse problem, characterized by challenges in fitting parameters with minor impacts, and occasionally necessitating the selection of more restricted models. Such issues have been noted by previous studies involving force constant fittings from phonon dispersions[9,10] and density of states[11]. These overfitting issues associated with optimization complicate comparative analysis across multiple spectra.

One prospective avenue to explore neutron scattering spectra is to employ machine learning techniques[12–21]. The integration of machine learning across various scientific disciplines is revolutionizing our approach to solving complex challenges, heralding a new era of discovery and innovation. In physical science, machine learning's capacity to navigate and interpret complex datasets is unlocking new frontiers, unveiling patterns and relationships that elude traditional analytical methodologies, offering innovative solutions to longstanding challenges. Such techniques have proven successful in addressing various problems, such as extracting magnetic interactions from diffuse neutron scattering[15–17] and retrieving structural parameters from polarized neutron reflectometry[18]. However, the scarcity of experimental data and the absence of known ground truth for continuous properties pose significant challenges. To tackle these issues, researchers resort to physics-based models to generate a substantial volume of labeled data for training their machine learning models. These models often encounter performance degradation when applied to experimental data due to the domain shift between physics-based simulations and real-world data. Therefore, it is imperative to thoroughly investigate this domain adaptation problem and develop effective strategies to mitigate its impact[20–23].

Here we present a machine learning-based framework for extracting force constants from two-dimensional INS measurements. Our approach employs a two-stage training design. The first stage involves self-supervised learning to create a feature extractor, encoding complex INS spectra into concise, meaningful latent representations that capture essential patterns and correlations. In this stage, a variational autoencoder[24] is trained using simulated spectra and subsequently fine-tuned with experimental spectra. The second stage leverages supervised learning, applying simulated data to train a regressor network capable of retrieving force constants from the latent representation. For inference, the experimental spectra of polycrystalline aluminum at different temperatures are processed through the feature extractor and regressor, and force constants was acquired. To demonstrate the advantages of our approach, we conducted a comprehensive comparison between our variational models, deterministic models, and a traditional optimization method. This framework is of great help in utilizing the two-dimensional $\chi''(|\boldsymbol{Q}|, E)$ data and would facilitate the exploration of the phonon dynamics of broader classes of materials efficiently. Furthermore, it has the potential to provide insights into a broader domain where researchers rely on physics-based simulations to understand experimental data.

## Results and Discussion

### Framework Details

Our approach to analyzing experimental INS spectra is tailored to navigate the inherent constraints of these data, notably their limited availability and the absence of labels. It is impractical to collect a large dataset of experimental INS spectra, as each experiment necessitates expert knowledge, experience of operating the instrument, and substantial time and investments. Meanwhile, assigning accurate labels of force constants is not only resource-intensive but also unreliable due to the ill-posed nature of this inverse problem. Such limitations present significant challenges for the direct implementation of end-to-end supervised models, as these methodologies typically require large, labeled datasets. To address this, we devised a two-stage framework that effectively harnesses the potential of both simulated and experimental spectra, as shown in Fig. 1.

In the initial phase of our framework, we adopted a self-supervised learning that bypasses the need for labeled data to construct a feature extractor. This component transforms the phonon information-rich yet unstructured spectra into a lower-dimensional latent space. By capturing the most relevant features while eliminating irrelevant information and noise, this process enables a more concise and informative representation that could serve as a foundation for subsequent tasks.

Autoencoder is a powerful deep learning technique with a bottleneck architecture, capable of compressing the input into a latent code through an encoder network and then decompressing the code into an output matching the dimension of the input through a decoder network. Minimizing the reconstruction error during training enables the networks to autonomously learn the optimal compression and decompression of the data. In our study, we implemented variational autoencoder, a special form of autoencoders, to transforms the input spectrum, comprising thousands of pixels, into a 30-dimensional latent code. Variational autoencoders are known for their robust data representation capabilities and are widely used in content generation applications. Unlike the traditional autoencoders that directly produce a single latent code, the variational autoencoders adopt a probabilistic approach and generate a 30-dimensional distribution to sample the latent code. The training process for variational models aims to minimize not only the reconstruction error but also the Kullback–Leibler (KL) divergence term, which quantifies the dissimilarity between the generated distribution and a prior distribution. This probabilistic approach encourages the latent space to exhibit a smoother and more structured representation of the input data. Although convolutional neural networks (CNN) are renowned for their effectiveness in various computer vision tasks[25–30], we employed fully connected (FC) layers as the backbone of networks. In fully connected neural network, each node from the previous layer is connected to each node of the current layer. Whereas in convolutional neural network, there are convolutional kernels that slide across the preceding layers, performing local operations. To assess the impact of incorporating variational components and fully connected layers, we conducted an extensive examination and evaluation of various types of autoencoders. The models examined in this study include fully connected autoencoder (FCAE), fully connected variational autoencoder (FCVAE), convolutional autoencoder (CNNAE), and convolutional variational autoencoder (CNNVAE).

The autoencoders were firstly trained on abundant simulated spectra. Although the simulated spectra may not perfectly replicate the complexities of real-world neutron data, they serve as a valuable substitute for training purposes, allowing us to establish a robust groundwork based on the physical models. The subsequent fine-tuning was carried out using experimental spectra with a reduced learning rate, serving to refine the model further to adapted to the nuances present in real-world data. Aluminum was selected as the material for framework verification due to its simple structure and extensive prior research. The simulation of aluminum spectra was performed with uniform random sampled force constants considering interactions up to the next-nearest neighbors. Phonon calculations were performed using Phonopy[31], and the corresponding spectra were generated using Oclimax[32]. Upon completion of the training of the autoencoders, the encoders are frozen and served as feature extractors, transforming raw spectra into high-quality latent representations.

With a refined feature extractor in place, the second phase leverages supervised learning to develop a regressor that retrieves force constants from the latent representations. This regressor network, consisting of fully connected neural layers, undergoes training solely with simulated data. Detailed description of the architecture and training specifics for our complete models are outlined in the Methods.

**Model Evaluation**

The evaluation of our final models presents a unique challenge due to the absence of ground-truth labels for experimental INS data. To ascertain the accuracy and reliability of our predictive framework, we employed two specific metrics. The first metric examines the discrepancy between input spectra and the spectra simulated using force constants predicted by the model. The second metric scrutinizes the physical plausibility of the temperature dependence in the predicted force constants, serving as a critical indicator of the model's reliability in reflecting underlying material behaviors. For comparison, we also extracted force constants using optimization methods. Given that spectra generation is the most time-consuming aspect, calculating gradients becomes impractical. Among gradient-free approaches, a direct search method called Nelder-Mead optimization[33] was selected instead of swarm intelligence methods like particle swarm optimization[34] owing to its speed and efficiency when provided with appropriate initial guesses.

The normalized mean square error (NMSE) for models with the feature extractor trained with different autoencoders was calculated and averaged across the experimental test set, as shown in Table 1. The relatively large values of NMSE for all models stem from various factors, such as irreproducible noise, limitations of forward physics-based models, and a reduction in machine learning model performance due to domain shift. The NMSE of the optimization method was regarded as an approximate lower limit within the constraints of existing experimental noise and physics-based model. Differences from this benchmark were calculated to focus only on domain-shift-induced performance degradation. For ablation purposes, we also tabulated the results of two alternate training strategies to discern the benefit from fine-tuning. One employed only simulated spectra for training the feature extractor, while the other involved extended training on the simulation-only feature extractor. This additional training was conducted under identical conditions to the fine-tuning but utilized the simulated training set.

The results elucidate marked disparities in performance among models equipped with different feature extractors. Models employing variational feature extractors, FCVAE and CNNVAE, perform similarly to the optimization method, whereas models with deterministic feature extractors, FCAE and CNNAE, underperform comparatively. Given that all models predict commendable force constants from the simulation test set, such divergence underscores their varying capability for applications in the experimental domain. Notably, feature extractors constructed with fully connected layers tend to surpass their convolutional counterparts. Moreover, fine-tuning with experimental spectra has proven to be effective in further enhancing the models' adaptability to the experimental domain for variational models. And this improvement is evidently linked to the integration of experimental data, rather than merely extending the training on simulated data.

Comparative analysis of the experimental spectra with those simulated from the predicted force constants was conducted by averaging the intensity along the Q-axis, as illustrated in Fig. 2. Two prominent peaks are discernible in the Q-averaged spectra, occurring at approximately 20 and 36 meV. With increasing temperature, a downward energy shift is observed for both peaks. This temperature-dependent trend is successfully captured by both the optimization method and our model that incorporates the FCVAE feature extractor. While the CNNVAE model accurately predicts the lower energy peak, it fails to replicate the behavior of the higher energy peak. Conversely, models utilizing FCAE and CNNAE feature extractors predict a trend for both peaks that is inverse to the one observed experimentally. Complementing this qualitative assessment, the NMSE values presented in Fig. 3(a) quantitatively substantiate the performance of the various models across different temperatures.

The validity of the predicted force constants was assessed by evaluating their temperature dependence, as illustrated in Figs. 3(b)-(f). Given the absence of any phase transition within the examined temperature range, the force constants are anticipated to exhibit a monotonic change. Consistent with this expectation and aligning with previous literatures[10,11], the variational models and the optimization method demonstrate a decreasing trend in the dominant force constant as temperature increases. In contrast, deterministic models depict an incongruent trend, suggesting a misalignment with physical realities. Regarding other subordinate force constants, it is observed that the optimization method does not demonstrate coherent, discernible temperature-dependent trends, and the deterministic models produce more variant results, possibly indicating their lack of stability and reliability. Conversely, our variational models, particularly the one utilizing the FCVAE feature extractor, give results with clear and physically reasonable temperature-dependent trends, lending credence to their reliability.

By integrating these two metrics, we have confirmed the effectiveness of our framework and the utilization of variational autoencoders with fully connected layers. Our findings suggest that models employing variational feature extractors exhibit less performance reduction across various domains compared to those employing deterministic feature extractors. Furthermore, performance can be enhanced by fine-tuning with target domain data, as enabled by our two-stage framework design. This makes our proposed framework valuable both before and after conducting neutron experiments. Models can be trained prior to the experiment using only simulated spectra, providing fast and accurate results for ad hoc decision making during the experiment. Once all experimental data is collected, the models can be further refined utilizing this real-world dataset to provide better results. Additionally, the performance of convolutional neural networks appears suboptimal, likely owing to the fact that global information is more crucial in spectral analysis than localized patterns that are typically sensitive to convolutional layers, as in the case of computer vision. Moreover, our variational models outperform both the optimization method and the deterministic models in delivering consistent and reliable predictions, particularly regarding temperature dependence. While the optimization method can generate force constants that perfectly mimic the experimental input spectra, the values may prove to be physically unrealistic. It is also important to note that the optimization process requires repetition for each test example, leading to increased time and resource demands as the number of examples grows. On the contrary, our machine learning method demonstrates high efficiency once training is completed, which is particularly valuable for real-time decision making during an experiment.

**Latent Space Analysis**

The pronounced performance difference between variational and deterministic models on experimental data prompted an in-depth investigation of the latent spaces they generate. We focused on autoencoders trained solely on simulation spectra, where the number of problem variables is known to be five. Figures 4(a)-(d) illustrate the latent codes produced by deterministic feature extractors and the latent distributions generated by variational feature extractors for simulation test set. Deterministic models encode input spectra into 30 active latent dimensions, all spanning similar distribution ranges. Conversely, variational models manifest a selective activation of latent dimensions, with only a subset being actively utilized and the remainder conforming to the prior normal distribution with a mean of zero and a standard deviation of one.

This difference in latent space activation was also analyzed by evaluating the sensitivity of decoder-reconstructed spectra to individual latent dimensions. The initial latent code was obtained from one of the test data, and mean values are used for variational models for simplicity. By individually varying each latent dimension while maintaining others constant, we isolated the effect of each dimension on the models' output. A variation of 0.5 was applied, yet similar results could be observed as long as the latent code remains within a reasonable range. The NMSE between the original and perturbated reconstructed spectra was calculated for each dimension and normalized to represent sensitivity. As depicted in Figs. 4(e)-(h), reconstructed spectra from deterministic autoencoders exhibited a uniform sensitivity across all latent dimensions, while the reconstructed spectra from variational autoencoders demonstrated sensitivity to only a few dimensions. Furthermore, Supplementary Figs. S2-S5 show how the reconstructed spectra of different autoencoders change when varying each dimension in the latent space, offering a visual representation of the effect of individual latent dimensions in the $(|\boldsymbol{Q}|, E)$-space.

Subsequent analysis, visualized in Figs. 4(i)-(l), explores the correlations among latent dimensions. For deterministic models, the Pearson correlation matrix reveals a high degree of correlation among several dimensions, as indicated by their elevated pairwise correlation coefficients. For variational models, the correlation matrix confirms the relative independence of their latent dimensions as the correlations between different dimensions are small. This behavior is attributed to the nature of variational models, as they also aim to minimize the KL divergence term in the loss. The presence of two dimensions representing identical features would introduce redundancy, thereby increasing the loss. As a result, the models tend to utilize a single dimension for each feature, ignoring redundant dimensions.

It is posited that these differences in latent space contribute to the models' divergent performance with experimental data. Deterministic models extract highly correlated latent codes that work collectively and contribute to the regression of force constants. This approach is effective for the simulation test set, as the training is specifically tailored to this type of data. However, when dealing with experimental data, which exhibit a domain shift relative to the simulation data, the intrinsic relations between these latent codes can be disrupted, hindering their collective functionality, and resulting in poor regression outcomes. In contrast, the variational models extract disentangled latent codes, each corresponding to distinct data patterns. This characteristic enhances the model's resilience to domain shift. Each latent dimension independently captures and represents relevant information, unaffected by the disrupted relationships evident in deterministic models, thus serving as a more robust foundation for regression.

**Model Generalizability**

To further highlight the advantages of our variational models, investigations were conducted to assess the impact of latent space dimensionality on their performance. Considering that the simulation training data was generated by altering five force constants, the selected setting of 30 latent dimensions should be sufficiently large. The dimensionality of the latent space was incrementally reduced from 30 to 3 to examine the resulting variations. As shown in Table 1, the performance of deterministic models showed marked fluctuations with changes in latent space dimensionality, sometimes resulting in exceptionally poor outcomes. Notably, aligning the number of latent dimensions closely with, or even reducing them below, the count of problem variables did not consistently enhance the performance of these deterministic models. On the contrary, variational models demonstrated robustness, maintaining lower NMSE as latent space dimensionality varies.

The resilience of variational models to changes in latent space dimensionality, attributable to their ability to extract disentangled representations, not only makes them suited for this specific problem, but also signifies their generalizability across various material systems beyond aluminum and at different experimental settings without concerns related to overfitting and hyperparameter tuning. This framework possesses the capability to discern and activate the requisite latent dimension automatically for each specific training dataset, illustrating their utility for tackling increasingly complex problems. Moreover, analysis of active latent dimensions within a dataset may facilitate identifying the correct dimensionality of the problem. A practical example would be to determine the number of force constants permitted by a set of experimental data with certain signal-to-noise ratios.

Building on this groundwork, the innovative two-stage framework we introduced holds remarkable potential beyond its initial scope, which focuses exclusively on a single task for a certain material. While the second stage of our framework is tailored for force constants regression, the first stage fulfills a more universal role in spectra understanding. The architecture's flexibility suggests the feasibility of further segregating the training dataset between these stages, enabling the development of a universal pre-trained feature extractor during the first stage. This would involve harnessing a diverse and extensive collection of unlabeled data from a variety of materials, including both experimental and rich simulated data. Such pre-trained feature extractor is anticipated to exhibit superior performance in general spectra understanding, establishing a robust foundation for downstream tasks. This approach may empower researchers to concentrate on refining task-specific models, like force constant regressors, utilizing smaller, targeted datasets. The implications of such a development are not limited to regression tasks but extend to classification, clustering, or generative tasks, promising new opportunities in materials science research.

## Conclusion

In summary, we have proposed a VAE-based framework for extracting force constants from experimental two-dimensional INS spectra. To the best of our knowledge, this is the first attempt to apply machine learning techniques for the comprehensive extraction of phonon dynamics from such spectra. Our investigations have highlighted a notable decrease in the performance of models trained on physics-based simulations when confronted with real-world data. We have shown that such a domain adaptation problem can be effectively addressed by our variational models. This improvement is attributed to the remarkable ability of the variational models to construct disentangled representations in the latent space. Moreover, our method demonstrates high efficiency in processing multiple test examples and excels at capturing consistent and reliable temperature-dependent trends, surpassing the limitations of the traditional optimization method. The inherent two-stage nature of our framework also makes it possible to fine-tune the feature extractor using experimental data, further reducing the performance degradation introduced by domain shift. The extensive analysis presented in this work not only proves the generalizability of our models in more complex systems, but also suggests the feasibility of developing a universal pre-trained feature extractor. The findings of this study provide valuable insights for future studies that aim to leverage machine learning techniques in the interpretation of complex spectra, especially in scenarios where the availability of experimental data is limited.

## Methods

**Details of models' architecture and training**

The architectures of autoencoder and regressor networks are schematically illustrated in Supplementary Fig. S1. The function of each layer and the final hyperparameters are discussed below.

The FC encoder network begins with a masked flatten layer, which flattens the pixels within the kinetic limit into 4144 neurons. In FCAE, two FC layers would down sample the input into 400 and then 30 neurons (latent code $z$), each followed by a tanh activation layer to introduce non-linearity. In FCVAE, the difference is that the second FC layer would output 30 normal distributions, represented by 30 mean neurons $\mu$ and 30 standard deviation neurons $\sigma$. The latent code $z$ of variational models will be sampled from $N(\mu, \sigma)$ at each time. The FC decoder takes latent codes as input, and up sample $z$ into 400 and then 4144 neurons by two FC layers, each followed by a tanh activation layer. The masked unflatten layer would turn the result into an image with the same size of the input spectrum.

The CNN encoder network begins with a cleaning layer to replace NaN values outside of the kinetic limit into 0s. The $70 \times 101$ spectrum is feed into 3 convolutional layers with 32, 64, 128 filters, with kernel size of $3 \times 3$ and stride of 2. After the convolutional layers and activation layers, the 128-channel $9 \times 13$ feature map is flattened and feed into two FC layers to down sample into 200 neurons and then 30 latent codes or distributions. The CNN decoder first up sample the latent codes into 200 and then 14976 neurons, which can be unflatten into a 128-channel $9 \times 13$ feature map. The feature map is then feed into 3 transposed convolution layers with 64, 32, 1 filter, with kernel size of $3 \times 3$ and stride of 2, each followed by an activation layer. After this, the region outside of kinetic limit of the $70 \times 101$ image will be replaced by NaN by a masking layer to output the final spectrum.

The autoencoder networks were trained using batch size of 64 for 2000 epochs. We utilized the stochastic gradient descent (SGD) as the optimizer with learning rate of 10 for FCAE and 1 for others and momentum of 0.9. The loss of the deterministic models is the mean square error between the input and output spectrum. In variational models, there is an additional Kullback–Leibler (KL) divergence loss that quantifies the dissimilarity between the current distribution and a prior distribution $N(\mathbf{0}, \mathbf{1})$. The weight of the KL divergence loss is set to 1e-6 after test experiments.

The regressor network contains 4 FC layers that followed by tanh layers and a final FC layer to output 5 force constants, where the first four FC layers transform the latent codes into 100, 1000, 1000, 100 neurons respectfully. The regressor was trained using batch size of 64 for 2000 epochs, with SGD optimizer with learning rate of 0.05 and momentum of 0.9. The loss is defined as the mean square error between the predicted force constants and the ground truth values.

Most of the training, inference, and optimization tasks were conducted on a single CPU core of an Intel Broadwell node, with only the CNN models being processed on a NVIDIA K80 GPU. Training times were approximately 6 hours for FC autoencoders and 4 hours for regressors, while CNN models required about 4.5 hours and 2.5 hours, respectively. Machine learning model inferences took only seconds. For comparison, the Nelder-Mead optimization, using reference values from ab initio calculation as the initial guess, took about 1.5 hours per test example.

**Experimental details**

Experimental data were obtained from polycrystalline aluminum samples using the time-of-flight instrument ARCS at the Spallation Neutron Source of Oak Ridge National Laboratory[35]. The aluminum samples were mounted inside a low-background electrical resistance vacuum furnace. The measurements were performed with an incident neutron energy of 50 meV, at 50, 150, 280, 350, 450, 540 and 640 K. Data reduction and analysis of the ARCS data were performed with MANTID[36]. The data were normalized by the proton current on the spallation target. Bad detector pixels were identified and masked, and the data were corrected for detector efficiency using a measurement of a vanadium standard. After data reduction, neutron events at different detectors were combined to generate the two-dimensional dynamical susceptibility $\chi''_{exp}(|\mathbf{Q}|, E)$.

**Simulations**

Given a set of force constants considering interactions up to the next-nearest neighbors, phonons were calculated using Phonopy. Corresponding two-dimensional spectra were then simulated using Oclimax with the same parameters employed in the experiment. A dataset of 10,000 simulated spectra was generated using force constants with each dimension randomly sampled from 0.5 to 1.5 of the reference value. The preparation of the dataset required approximately two days. Considering that we utilized only a single CPU core, reducing the processing time could be easily achieved by allocating more resources. The dataset was divided into three subsets: 80% for training, 10% for validation, and 10% for testing purposes.

**Data preprocessing**

All the spectra were cropped to 0 - 10 Å and 10 - 45 meV, focusing on the relevant phonon region. Subsequently, they were resized to the same resolution. The background was estimated as a function of energy and subtracted from the experimental data to eliminate unwanted noise. Finally, mean normalization was applied to each spectrum, ensuring a meaningful comparison between the simulated and experimental spectra with varying scan times.

**Acknowledgements**

This material is based on work supported by the U.S. Department of Energy, Office of Science, Office of Basic Energy Sciences, Neutron Scattering Program under Award Number DE-SC0023874. The neutron scattering portion of this research used resources at the Spallation Neutron Source, which is a DOE Office of Science User Facilities operated by the Oak Ridge National Laboratory.

**Author Contributions**

Y.S. and C.L. conceived the project. Y.S. generated the simulated dataset, preprocessed experimental data, and applied machine learning techniques. All authors contributed to the data analysis, result interpretation, and manuscript writing.

**Data availability**

The data that support the findings of this study are available at https://doi.org/10.5281/zenodo.10373288[37].

**Code availability**

The codes developed in this study are available at https://github.com/suyaokun/2DPhononINS-ML.

**Competing Interests statement**

The authors declare no competing interests.

# Figures and Tables

**Fig. 1. Framework architecture.**

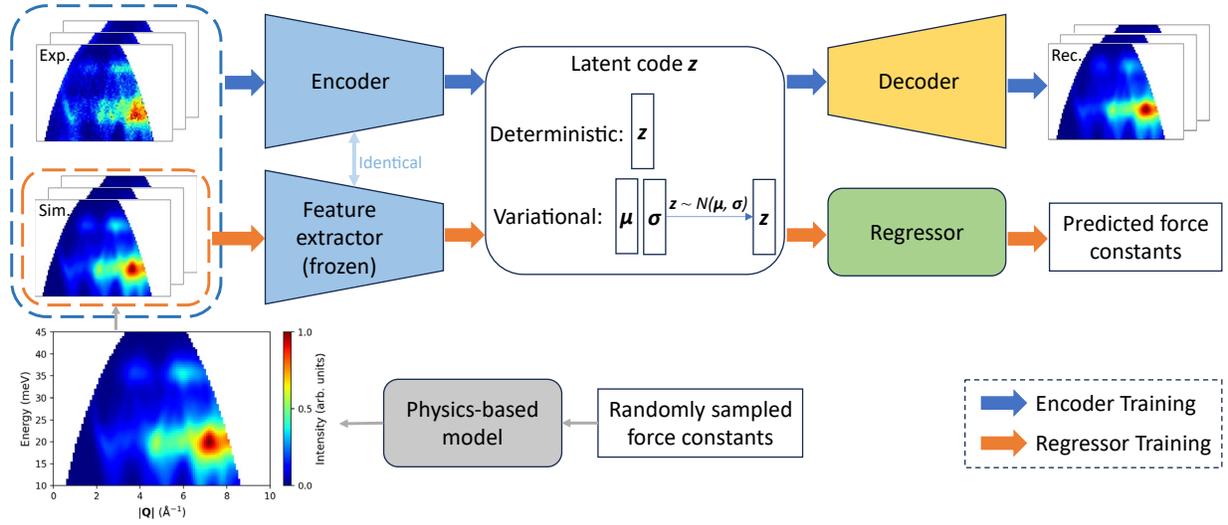

**Fig. 1. Framework architecture.** Our two-stage framework integrates self-supervise learning and supervised learning for phonon dynamics analysis from INS spectra. In the first stage, an autoencoder is trained to transform the input spectrum into a latent code through an encoder network and then reconstruct the spectrum back via a decoder network. Upon completion, the encoder is frozen and served as the feature extractor for the subsequent stage. The training utilizes simulated spectra, generated from physics-based model with randomly sampled force constants, and is fine-tuned using experimental spectra to adapted to real-world data. While deterministic models directly produce a single latent code, variational ones adopt a probabilistic approach and generate a distribution to sample the latent code. In the second stage, a regressor network, trained on labeled simulated spectra, is used to predict force constants.

**Fig. 2. Comparison between the experimental spectra and simulated spectra with the model predicted force constants at varying temperatures.**

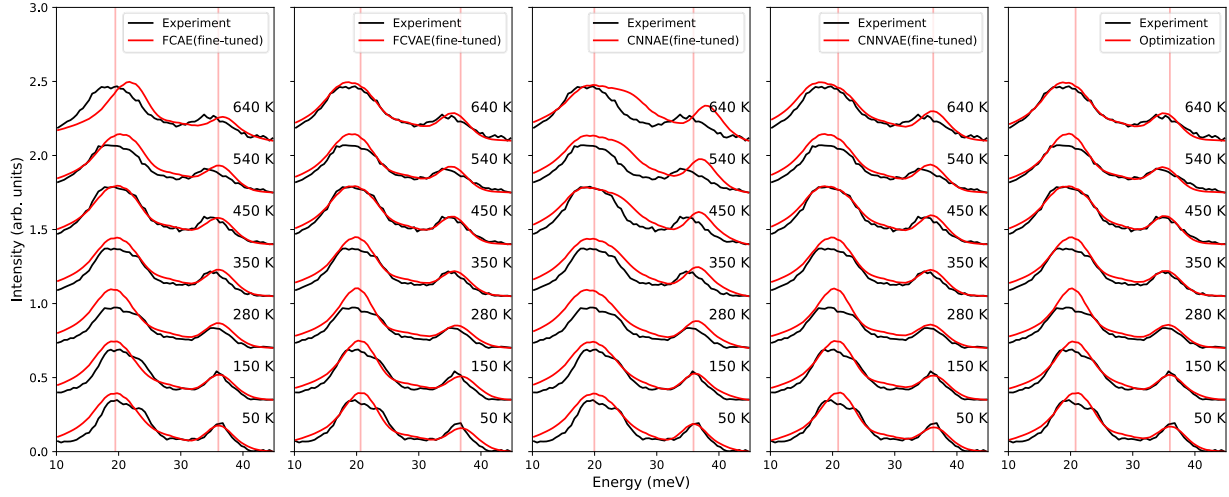

**Fig. 2. Comparison between the experimental spectra and simulated spectra with the model predicted force constants at varying temperatures.** The intensity curves are obtained by averaging over the Q-axis of the spectra at different temperatures and offset for ease of comparison. Both the optimization method and our FCVAE model accurately predict the decreasing trend of two peaks, whereas the CNNVAE model succeeds only with the lower energy peak, and the FCAE and CNNAE models fail for both peaks. The red vertical stripes denote the peak positions in red curves at 50 K, offering a reference for observing peak position shifts with temperature.

**Fig. 3. Comparison of models' performance on experimental data.**

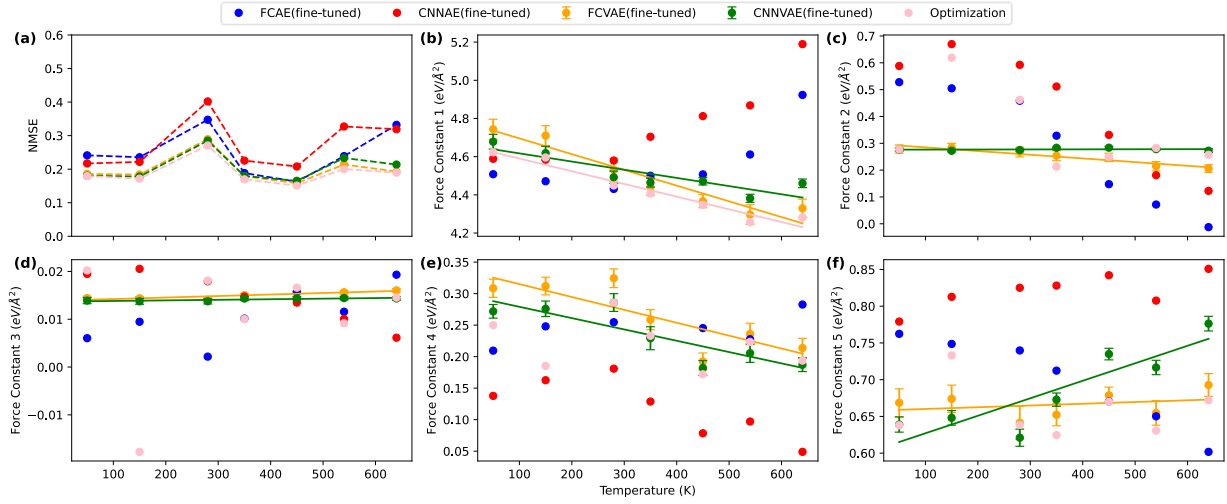

**Fig. 3. Comparison of models' performance on experimental data. a** NMSE of different models and optimization method. Dashed lines are guides to the eye. **b-f** Force constants predicted by different models. For variational models, 10,000 trials were performed, with error bars representing the standard deviation of the outcomes. Solid lines are linear fit for results with clear temperature dependent trends.

**Fig. 4. Latent space analysis for models.**

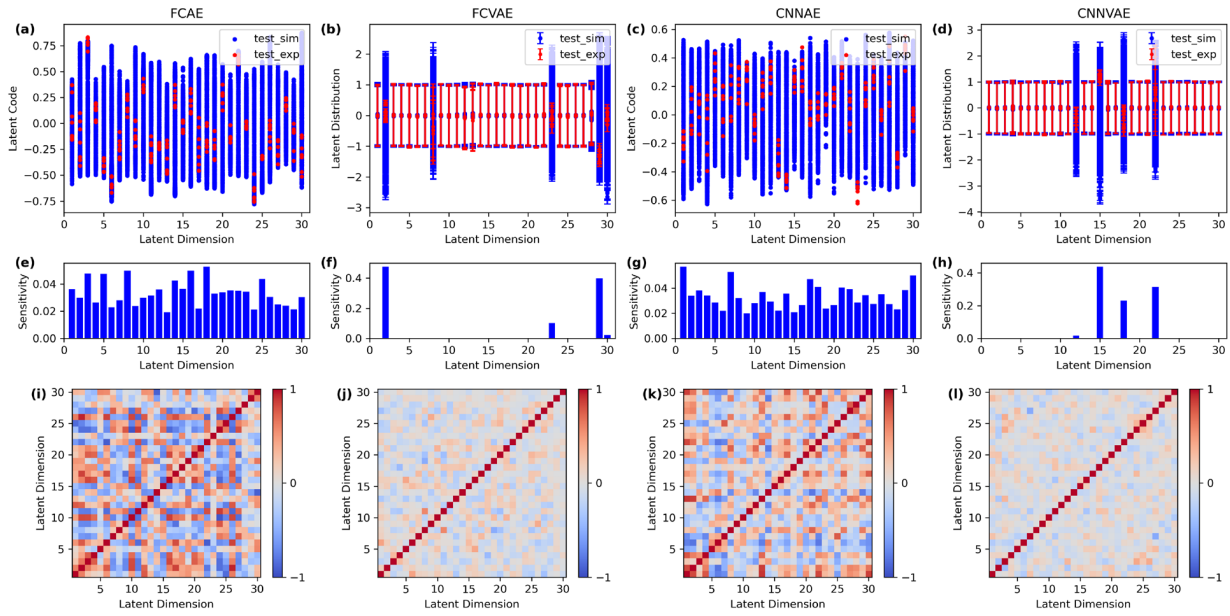

**Fig. 4. Latent space analysis for models. a, c** Latent codes of simulation test data (blue) and experimental data (red) for FCAE and CNNAE. **b, d** Latent distribution of simulation test data (blue) and experimental data (red) for FCVAE and CNNVAE. The circles and error bars represent the mean values and the standard deviations of normal distributions. **e-h** Sensitivity of decoder-reconstructed spectra to individual latent dimensions. **i-l** Pearson correlation matrix between different latent dimensions of simulation test set. For variational models, the correlations were calculated using the mean values of each dimension.

**Table 1. Averaged NMSE between experimental input spectra and the spectra simulated using force constants predicted by models.**

|  | Latent space dimension | FCAE | FCVAE | CNNAE | CNNVAE |
|---|---|---|---|---|---|
| Sim-only | 30 | 0.236 (0.046) | 0.201 (0.011) | 0.277 (0.086) | 0.212 (0.021) |
| Fine-tuned | 30 | 0.249 (0.059) | 0.200 (0.010) | 0.274 (0.084) | 0.205 (0.015) |
| Extended | 30 | 0.256 (0.066) | 0.202 (0.011) | 0.296 (0.105) | 0.211 (0.020) |
| Sim-only | 25 | 0.307 (0.117) | 0.208 (0.018) | 0.242 (0.052) | 0.206 (0.016) |
|  | 20 | 0.223 (0.033) | 0.205 (0.014) | 0.380 (0.190) | 0.213 (0.023) |
|  | 15 | 0.208 (0.017) | 0.206 (0.016) | 0.247 (0.057) | 0.208 (0.018) |
|  | 10 | 0.347 (0.157) | 0.209 (0.019) | 0.231 (0.041) | 0.200 (0.009) |
|  | 5 | 0.370 (0.180) | 0.206 (0.016) | 0.259 (0.069) | 0.211 (0.020) |
|  | 4 | 0.205 (0.015) | 0.207 (0.017) | 0.218 (0.027) | 0.223 (0.033) |
|  | 3 | 0.213 (0.022) | 0.205 (0.014) | 0.242 (0.052) | 0.204 (0.014) |

**Table 1. Averaged NMSE between experimental input spectra and the spectra simulated using force constants predicted by models.** The NMSE is calculated for models with the feature extractor trained with different autoencoders, each in three training stages: simulated spectra only (sim-only), fine-tuning with experimental spectra (fine-tuned), and extended training on simulated spectra (extended). The benchmark NMSE for Nelder-Mead optimization is 0.190. Differences from this benchmark are indicated in parentheses for each model.

# Supplementary Information:

# Uncovering Obscured Phonon Dynamics from Powder Inelastic Neutron Scattering using Machine Learning

Yaokun Su[1] and Chen Li[1,2*]

1. *Materials Science and Engineering, University of California, Riverside, Riverside, CA, United States.*
2. *Mechanical Engineering, University of California, Riverside, Riverside, CA, United States.*
    *chenli@ucr.edu

**Architecture diagrams of models**

The detailed architectures of autoencoder and regressor network are shown in Fig. S1. The functions of each layer and the final hyperparameters are discussed in Methods of the main text.

**Effects of individual latent space dimensions on reconstructed spectra**

To better understand the effect of individual latent dimensions, a comprehensive analysis was conducted to investigate the change of decoder-reconstructed spectrum when varying each dimension in the latent space. The procedure is the same as the sensitivity calculation in main text. Output spectra were reconstructed for each variation and compared with the one decoded from original latent code.

Figures S2-S5 illustrate the changes in reconstructed spectra corresponding to each varied dimension across four models. In deterministic models (Figs. S2 and S4), alterations in each dimension lead to similar amplitude changes in the output spectra, suggesting a relatively equal contribution from each dimension. In contrast, the variational models (Figs. S3 and S5) exhibit pronounced effects from specific dimensions, while influences from others are negligible. This suggests that these few dimensions are the main contributor in the models.

Combined with the results reported in the main text, these findings imply different approaches adopted by the models and their corresponding outcomes. The deterministic models utilize all dimensions in the latent space to represent the input spectra. While this tailored representation works effectively for the simulation test set, the cooperation among different dimensions becomes compromised with experimental data. On the other hand, variational models rely on a few key dimensions, each representing distinct features of the input spectra. Although their performance with training data is marginally inferior, they exhibit robustness and superior generalization in handling diverse or shifting domains.

# Figures and Tables

## Fig. S1

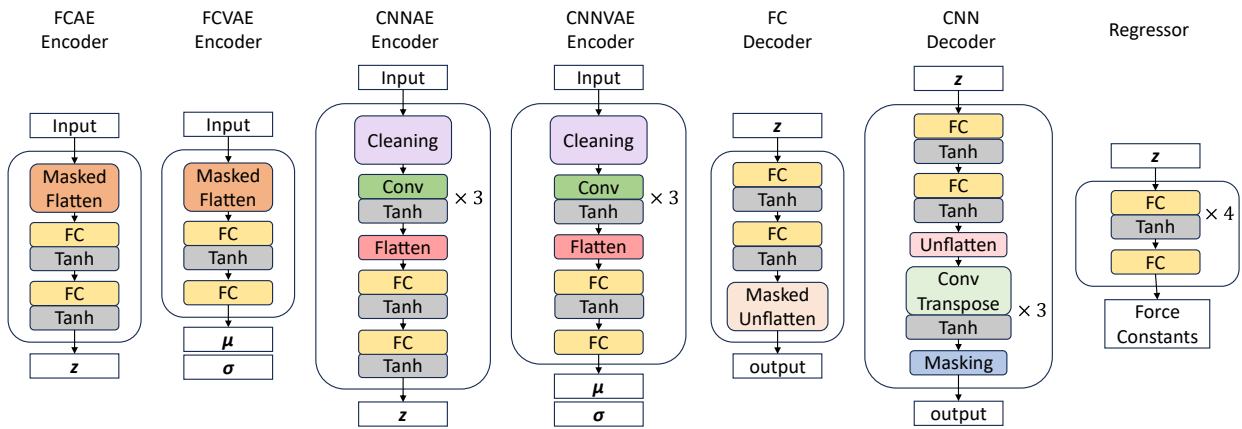

**Fig. S1. Architectures of Autoencoder and regressor networks.**

**Fig. S2**

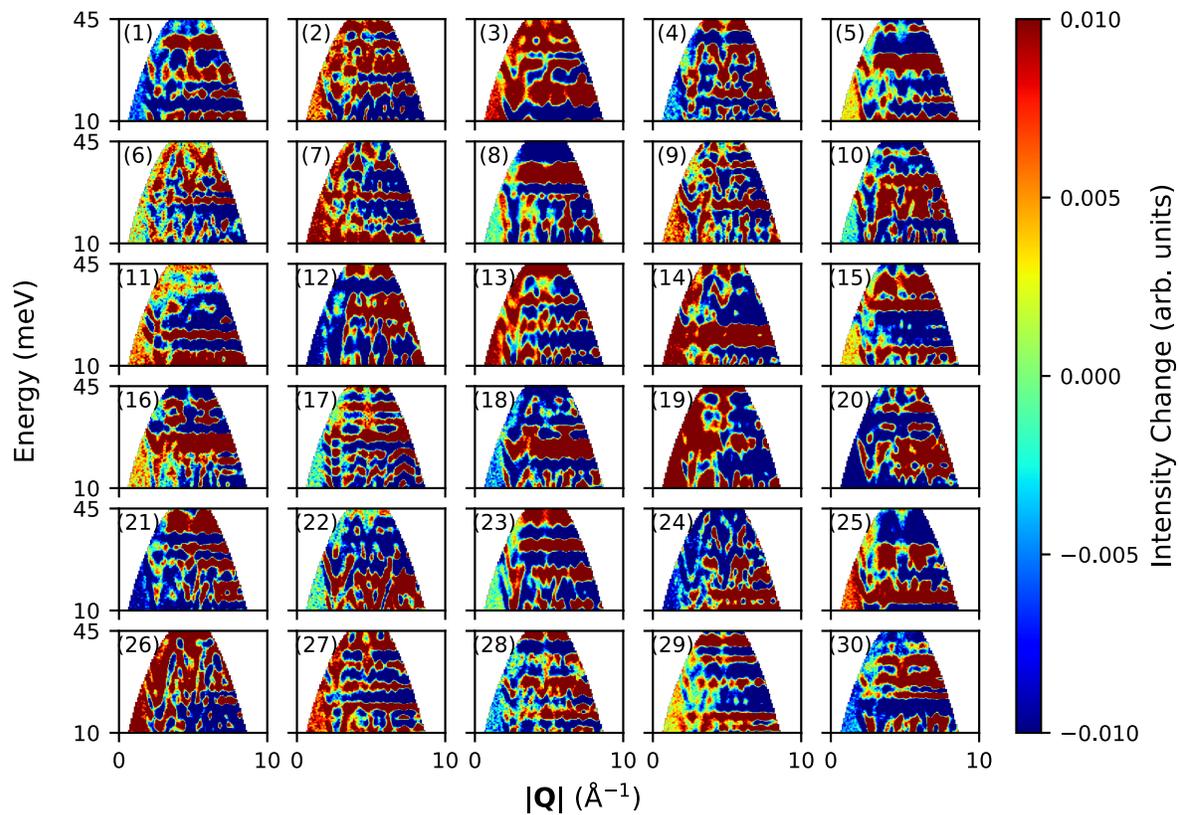

**Fig. S2. The change of FCAE decoder-reconstructed spectrum induced by variation in each latent dimension.** The numerical labels represent indices of dimensions in the latent space.

**Fig. S3**

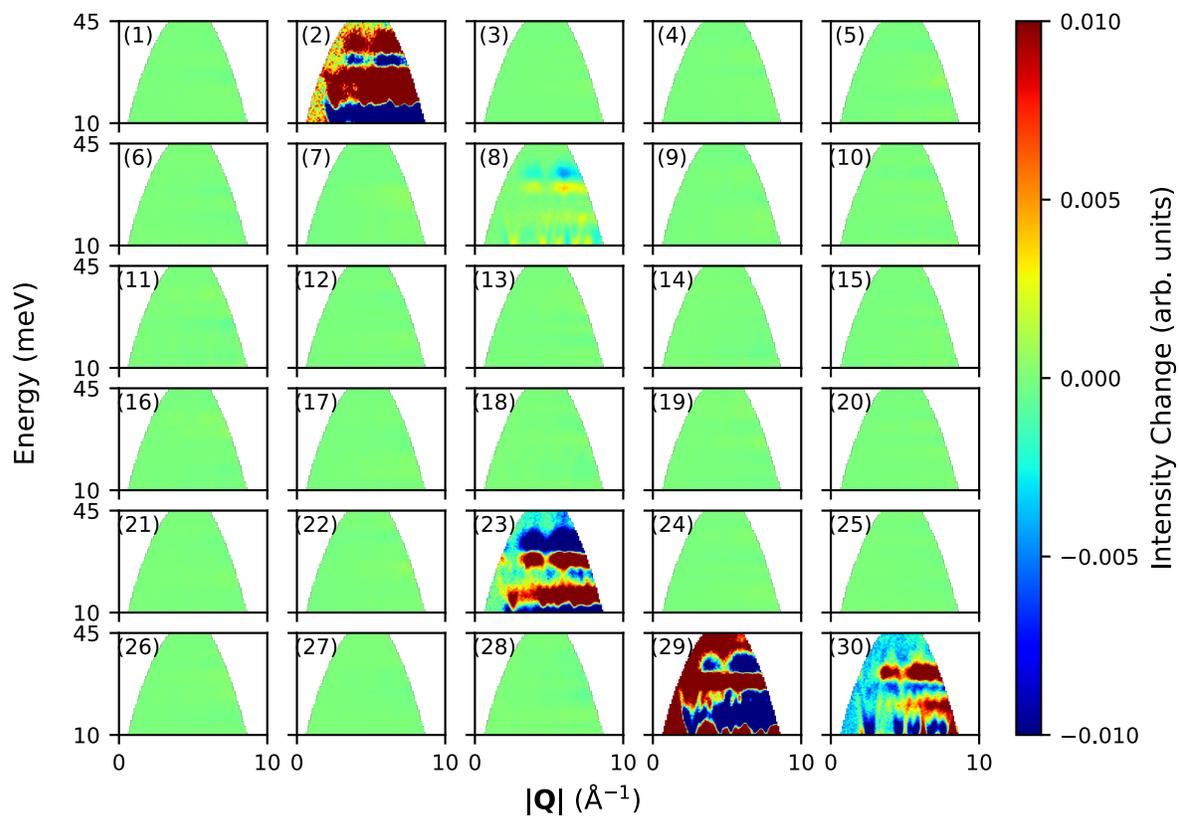

Fig. S3. The change of FCVAE decoder-reconstructed spectrum induced by variation in each latent dimension.

**Fig. S4**

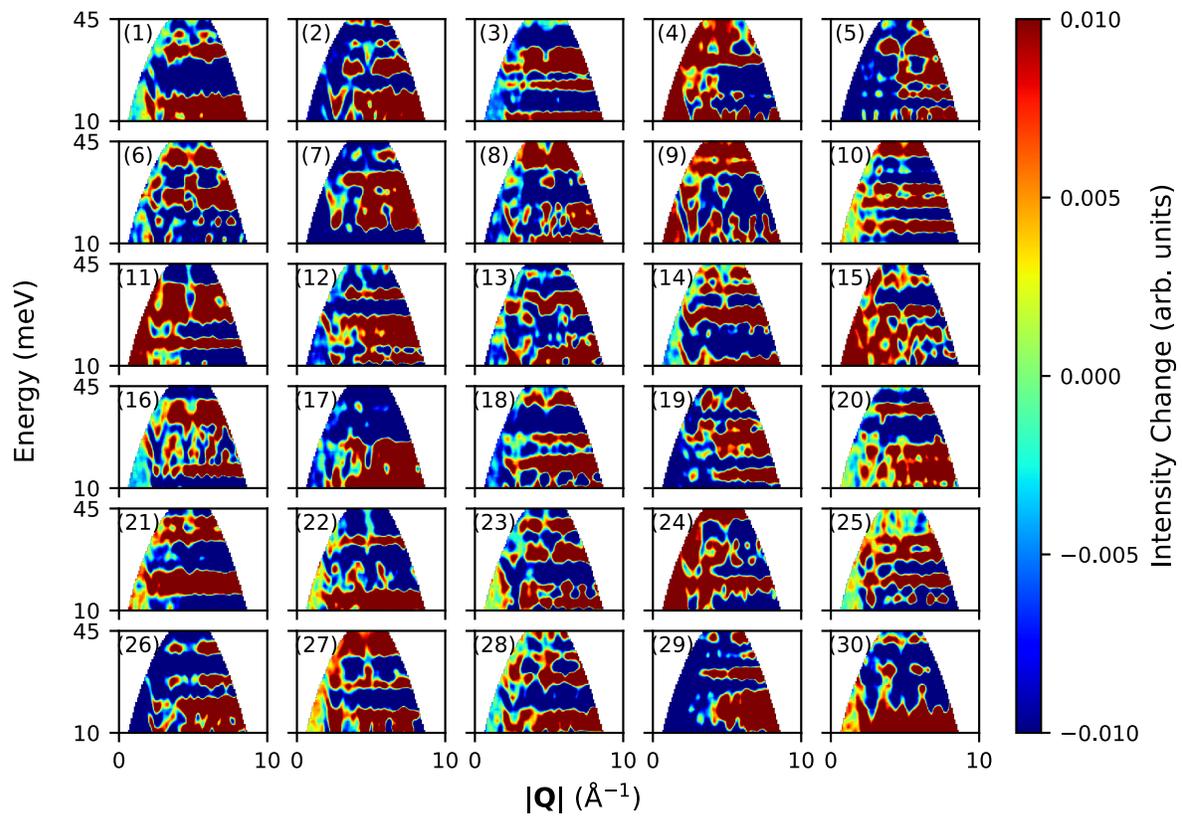

**Fig. S4. The change of CNNAE decoder-reconstructed spectrum induced by variation in each latent dimension.**

**Fig. S5**

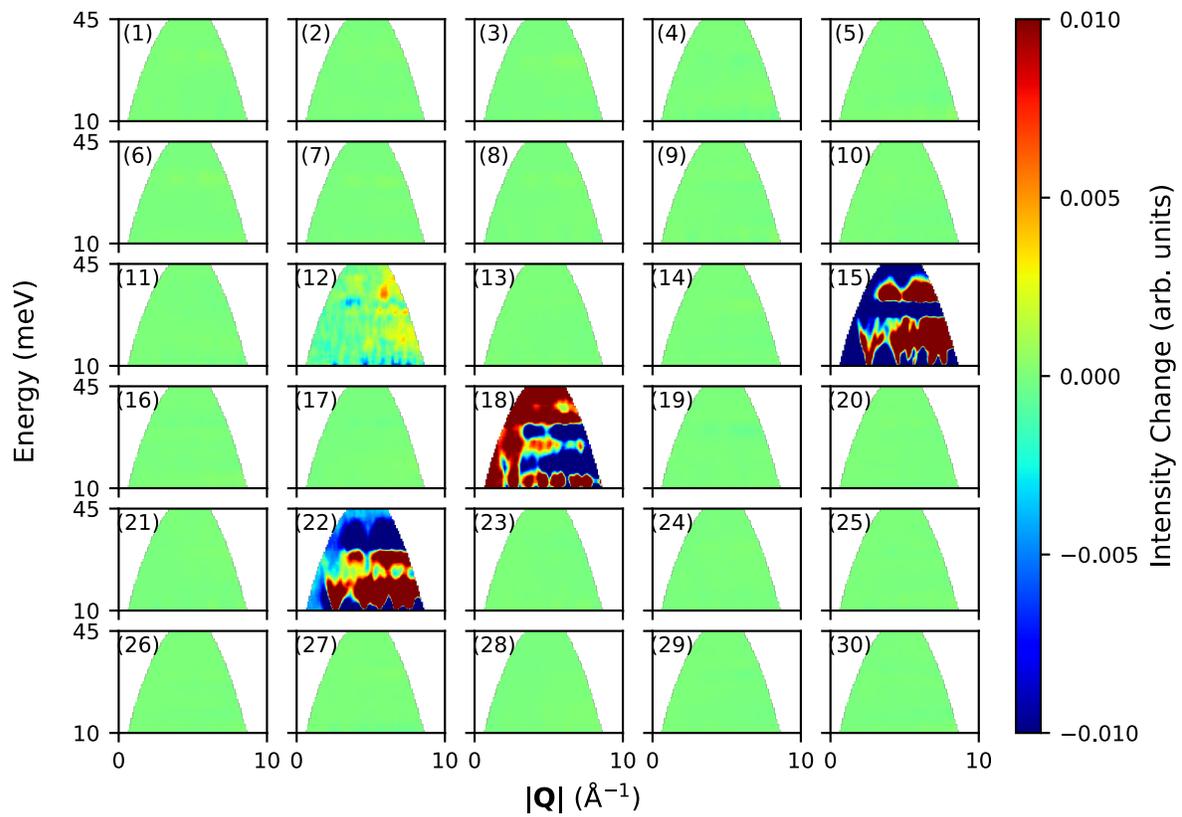

**Fig. S5.** The change of CNNVAE decoder-reconstructed spectrum induced by variation in each latent dimension.